\documentclass[fleqn,usenatbib]{mnras}

\usepackage{newtxtext,newtxmath}

\usepackage[T1]{fontenc}
\usepackage{ae,aecompl}

\usepackage{graphicx}	
\usepackage{amsmath}	
\usepackage{amssymb}	

\title[BZ Process in Long and Short GRBs]{Constraints on Gamma-ray Burst Inner Engines in a Blandford-Znajek Framework}

\author[Lloyd-Ronning et al.]{Nicole M. Lloyd-Ronning,$^{1,2}$\thanks{E-mail: lloyd-ronning@lanl.gov}
Chris Fryer,$^{1}$
Jonah M. Miller,$^{1}$
Neelima Prasad,$^{3}$
\newauthor Chris Torres,$^{2}$  
Phillip Martin$^{4}$
\\
$^{1}$Center for Theoretical Astrophysics and CCS-2, Los Alamos National Lab, Los Alamos, NM, USA\\
$^{2}$University of New Mexico, Los Alamos, NM USA\\
$^{3}$University of California, San Diego, CA USA\\
$^{4}$Cornell University, Ithaca, NY USA\\
}

\date{Accepted XXX. Received YYY; in original form ZZZ}

\pubyear{2018}

\begin{document}
\label{firstpage}
\pagerange{\pageref{firstpage}--\pageref{lastpage}}
\maketitle

\begin{abstract}
 Under the assumption that a Gamma-ray Burst (GRB) is ultimately produced by a Blandford-Znajek (BZ) jet from a highly spinning black hole (BH), we put limits on the magnetic field and BH mass needed to power observed long and short GRBs.  For BHs in the range of $2-10 M_{\odot}$ (for long GRBs) and $0.5-4 M_{\odot}$ (for short GRBs), we find magnetic fields in the range of $5x10^{14} \lesssim B \lesssim 10^{17}$ G are needed to power the observed GRBs. Under the simple assumption of flux conservation, we estimate the magnetic fields of the progenitor systems for both long and short GRBs, finding that single massive star progenitors require fields $\sim 10^{6}$ G and NS merger systems require fields $\sim 10^{15}$ G. We also discuss the implications and consequences of high magnetic fields in GRB BH-disk systems, in terms of MRI field growth and magnetically arrested disks.  Finally, we examine the conditions under which the progenitor systems can retain enough angular momentum to create BHs spinning rapidly enough to power BZ jets.
\end{abstract}




\section{Introduction}
 The progenitor system that gives rise to a gamma-ray burst (GRB) has long been an unresolved problem, but a general framework has emerged. Long GRBs appear to be associated with massive stars and short GRBs with the merger of two compact objects (reviews that compile and discuss the evidence behind these associations for both long and short bursts include \cite{pir04, ZM04, Mesz06, GRRF09,KZ15,Lev16}; for short bursts specifically, see \cite{LRR07,Berg14,DAvanz15}). 

  A black-hole (BH) accretion disk system is a plausible outcome of these cataclysmic events \citep{Woos93,MW99,Heg03,FM13,Fry15}, although it is possible that these progenitors produce a highly magnetized ($B \gtrsim 10^{14}$ Gauss) and potentially hypermassive ($M \gtrsim 2M_{\odot}$) neutron star (NS) after the collapse/merger \citep{Usov92,DT92,Thomp94,ZM01,Mazz14,Row14,Metz15,Rea15,SD18}.  The BH-disk system has been shown to be a viable engine for the GRB jet, from the standpoint of the timescales and energetics involved (e.g., consider the energy available from a $20 M_{\odot}$ star accreting $5 M_{\odot}$ at a rate of $.01 - 0.1 M_{\odot}s^{-1}$; for more discussion on this, see, for example, \cite{PWF99}).  
Observationally, we see BH-accretion disks in active galactic nuclei (which can, in many ways, can be considered scaled cousins of GRBs; \cite{Nem12,Zhang13,wu16,LR18}), and infer they are behind the relativistic jets observed in these objects. 

 It was suggested in the 1970's \citep{bz77} that a magnetized BH-accretion disk system is capable of launching a powerful jet through the so-called Blandford-Znajek (BZ) process.  The energy reservoir for the BZ process is the rotational energy of the central BH, extracted as the BH rotates in an external magnetic field. The process can be envisioned in a number of ways - for example, in terms of the torque exerted by the BH on magnetic flux tubes threading the horizon, and/or in terms of a dc-circuit analogy (a current is induced by changing magnetic flux near the horizon, and both the BH and force-free region around it have some effective impedance. One can calculate the power transmitted to the force-free region via Ohm's law).  We refer the reader to the  discussions in \cite{MT82} for a detailed analysis of both of these formulations.\footnote{The BZ process can also be modeled as a Penrose process. See \cite{BCP15} for a review and introductory discussion of Penrose processes and their wave analogues.}
 
  No matter how one formulates the problem, the process reduces to the essential point that - given some extant magnetic flux around the BH - frame dragging effects near the BH horizon cause field lines closer to the BH to rotate faster than those further from the BH.  The consequence of this can be (particularly if the angular velocity of the magnetic flux tubes rotate at approximately half the rate of the BH; \cite{MT82}) a powerful Poynting flux generated along the BH spin axis.
 
  In addition, recent simulations have shown that BZ jets can realistically be launched in spinning BH-accretion disk systems \citep{LWB00, mg04, Nag11,  tch11, McK12,Pen13, Brom16, Par18}), including those resulting from double NS mergers \citep{S17, Ruiz18}.  The efficiency of this process in terms of extraction of the BH spin energy and/or accretion rate is discussed in \cite{mg04,mck05,tch11}.   These simulations seem to indicate that the framework of the BZ mechanism powering a relativistic GRB jet is feasible (for additional detailed discussion of BZ jets in GRBs, see \cite{LWB00,LRR02,tm12,Lei13,Liu15,Lei17,xie17}; recent simulations and discussion of GRB jet launch in general can be found in \cite{Wang08,Mor10,Naga11,Miz13, LC13, LC16,Ito15,Harr18}).
  
  A successful jet of course needs the power to make its way through any material surrounding the poles. For long GRBs in particular, we expect some amount of material (from a few to $\sim 10 M_{\odot}$) from the massive stellar collapse surrounding the inner engine (a so-called cocoon \cite{RR02,Laz05,Mors07}).  In what follows, we assume GRB emission comes from a successfully launched relativistic jet - i.e. a jet that has indeed made its way through any surrounding material, and has only deposited a small fraction of its energy in the cocoon (see \cite{LML15} and references therein for a discussion of the issues involved in a successful jet launch).  
  
  Under the assumption that both long and short GRBs are powered by a successfully launched BZ jet, we constrain the necessary magnetic field and BH mass, assuming a rapidly rotating BH. We also explore how these constraints affect our progenitor models in terms of what we can say about the progenitor system's magnetic field and angular momentum.  
  
  Our paper is organized as follows: In \S 2, we discuss the theoretical framework of a BZ jet and the assumptions we make in our calculations. In \S 3, we present our results, exploring constraints put on the magnetic field and BH mass of long and short GRBs. We also make an estimate of the magnetic field of the progenitor system assuming magnetic flux conservation. In \S 4, we discuss magnetic field growth in GRB disks and potential implications for producing a magnetically arrested disk (MAD) scenario. In \S 5, we explore in more detail the necessary conditions for progenitors of GRBs to produce the required angular momentum to launch a BZ jet.  Conclusions are given in \S 8.

\begin{figure}
\centering
\includegraphics[width=\columnwidth,height=2.8in]{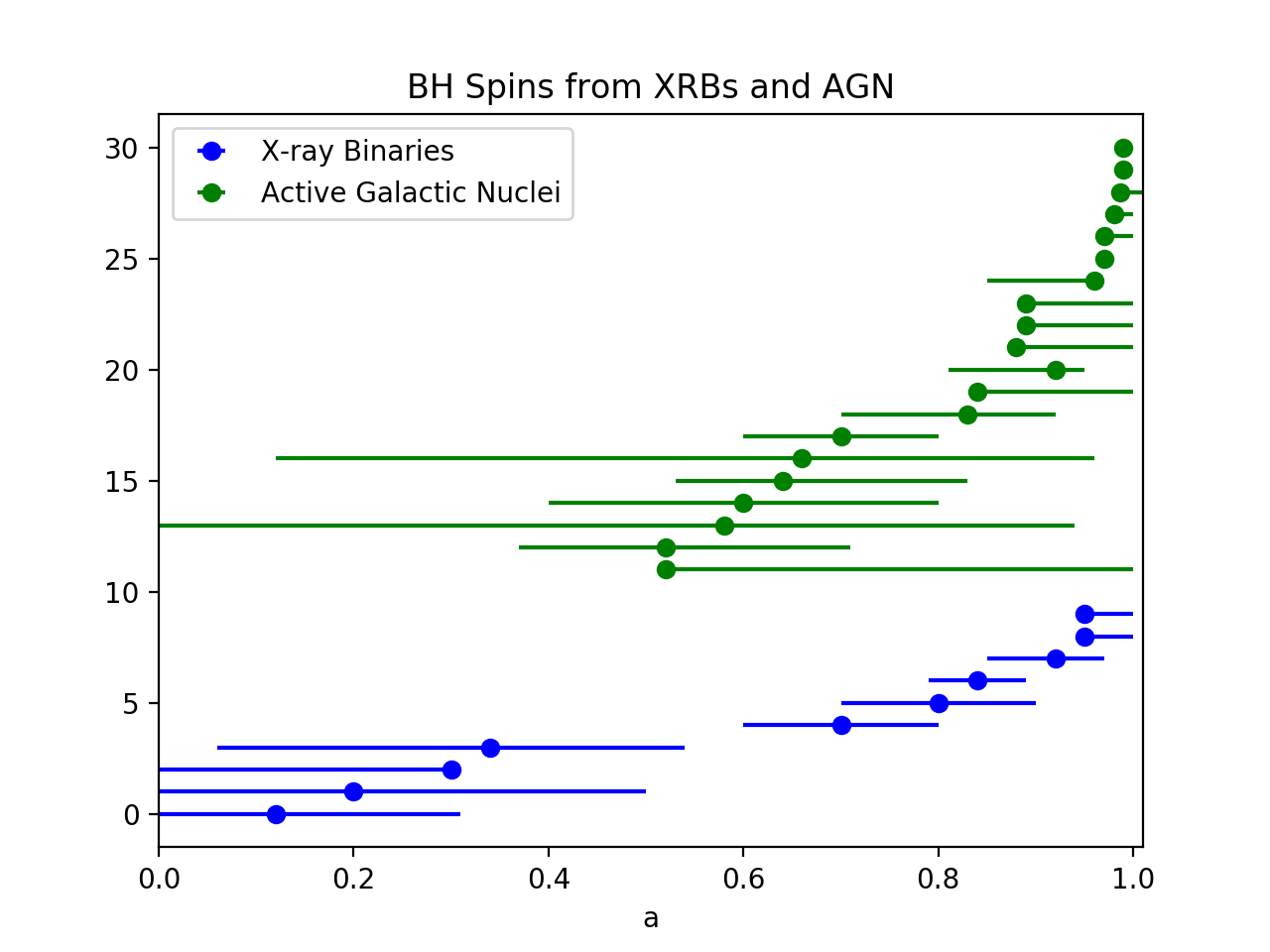} 
\caption{BH spin parameters inferred from 10 X-ray binaries (blue dots), and 20 AGN (green dots).  Data was taken from tables in McClintock et al., 2014, for the X-ray binaries, and Reynolds, 2014, for the AGN.}
\label{fig:bhspins}
\end{figure}

\section{Theoretical Framework and Assumptions}
  The analysis below is carried out in the context that the GRB jet is a result of the BZ process.  We can distill the details of the BZ process into a relatively simple expression for the luminosity from the jet (see, e.g., \cite{TNM10,TG15}):

\begin{equation}
\centering
\begin{aligned}
L_{BZ} = & (kfc^{5}/64\pi G^{2})a^{2}\phi_{BH}^{2}M_{BH}^{-2} \\ \\
      & \approx 10^{53} {\rm erg} \ (a/0.9)^{2}(\phi_{BH}/10^{29}{\rm G}cm^{2})^{2}(5M_{\odot}/M_{BH})^{2}
\end{aligned}
\end{equation} 

\noindent where k is a geometrical factor related to the magnetic field geometry (of order $\sim 0.05$), and $f$ is a correction factor related to the BH spin $a = Jc/GM_{BH}^{2}$ ($J$ being the BH angular momentum). For $a \lesssim 0.95$, $f \approx 1$. The parameter $\phi_{BH}$ is the magnetic flux near the horizon, and $M_{BH}$ is the mass of the BH.  Again, although this equation was originally derived for slowly spinning ($a \sim 0.1$) BHs, it has been shown to describe the BZ luminosity from more rapidly spinning BHs (up to $a \sim 0.95$) within the correction factor $f$ (for additional discussion of the validity of equation 1, see \cite{mck05, TNM10,TG15}).

  
 If the GRB ultimately derives its power from the rotational energy of the central BH via the BZ jet, the observed luminosity of the GRB, $L_{GRB}$, is some fraction $\eta$ of $L_{BZ}$: $L_{GRB} = \eta L_{BZ}$.
Under this basic premise, we can use the observed GRB prompt luminosity $L_{GRB}$ to explore the parameter space for which this equation is viable.
  
 Throughout our paper, we assume an efficiency factor $\eta$ of $0.1$ - i.e. $10$ percent of of the Blanford-Znajek jet power goes in to producing the gamma-ray luminosity.  The efficiency of conversion of jet energy to GRB radiation has been discussed extensively in the literature (see, e.g., \cite{guet01,lrz04,fp06}). Although we do not have a definitive handle on this number $\eta$ either observationally or theoretically, our value of $0.1$ is consistent with (and even conservative compared to) estimates from GRB afterglow observations \citep{lrz04,fp06,rp09}.  Note that our results below for the values of the magnetic field and BH masses will scale accordingly, changing by a factor of about 3 for every order of magnitude change in the efficiency $\eta$.   
 
  We assume a jet opening angle of $\theta \approx 10^{\circ}$, so that the luminosity in the GRB jet is $L_{GRB} = \frac{1}{2}(1-{\rm{cos}}(\theta))L_{iso}$, where $L_{iso}$ is the isotropic equivalent luminosity.  Although - again - we don't have extremely strong constraints on the jet opening angle from observations of long or short GRBs, in a number a cases a clear break in the afterglow light curve suggests the presence of jet with opening angles of $\sim 5$ to $10$ degrees (for example, see results reported in \cite{Ghir04,Berg14,fong15,Dain17,Wang18}). In some cases, the inferred opening angle can be less by a factor of 2 or so \citep{Wang18}. We point out the inferred average opening angle for long GRBs seems to be somewhat larger than that of short GRBs, although a value of $\theta = 10^{\circ}$ lies in the range for both populations (see, e.g. Figure 18 of \cite{Berg14}). A larger jet opening angle will require more power from our BZ jet; hence, we consider $\theta \sim 10^{\circ}$ , a conservative estimate in the sense that it requires more extreme values of magnetic fields and BH masses compared to smaller jet opening angles.   \\
  
  We also assume a roughly poloidal magnetic field configuration on the BH, which produces the strongest possible jet (the details of the field structure are contained in the factor $k$ in equation 1 above). The magnetic field is estimated from the magnetic flux through the relationship $\phi \approx B \Omega R^{2}$, where $B$ is the magnetic field, $\Omega$ is the solid angle of the jet,  and $R$ is the Kerr radius given by $R = GM/c^{2} + \sqrt{(GM/c^{2})^{2} - a^{2}}$.  Note that expressing magnetic flux in terms of magnetic field and relating the Kerr radius to the BH mass, we see that:
 \begin{equation}
\centering
L_{GRB} \approx 10^{50} {\rm erg} (\eta/0.1) (a/0.9)^{2}(B/10^{16}{\rm G})^{2}(M_{BH}/5M_{\odot})^{2}
\end{equation}

  Throughout this paper, we assume an $a=0.9$, which is consistent (and in some cases comfortably above) the value needed to launch a GRB jet according to the simulations referenced in the introduction (e.g. \cite{,McK12, Ruiz18}).  Such high values of BH spin have indeed been inferred through observations of stellar mass BHs in X-ray binaries, as well as in supermassive BHs powering AGN (for reviews on these measurements, see \cite{MNS14} and \cite{Rey14}).  Figure~\ref{fig:bhspins} shows BH spin parameters from X-ray binary observations (blue circles) and AGNs (green circles), using data from the previous two references. The data indicate that the majority of these objects have high values BH spins $a \gtrsim 0.6$.
  On the other hand, estimates of the spins of remnant BHs from BH-BH merger seen by LIGO indicate very low ($a \sim 0.05$) inferred spin values. Hence, the distribution of BH spin parameter and its dependence on the progenitor system is still an open question.  We discuss the implications of our relatively high spin constraint - particularly in terms of the requirements it puts on GRB progenitor angular momentum - in \S 5.2 below.   

\begin{figure*}
\centering
\includegraphics[width=3.5in,height=2.7in]{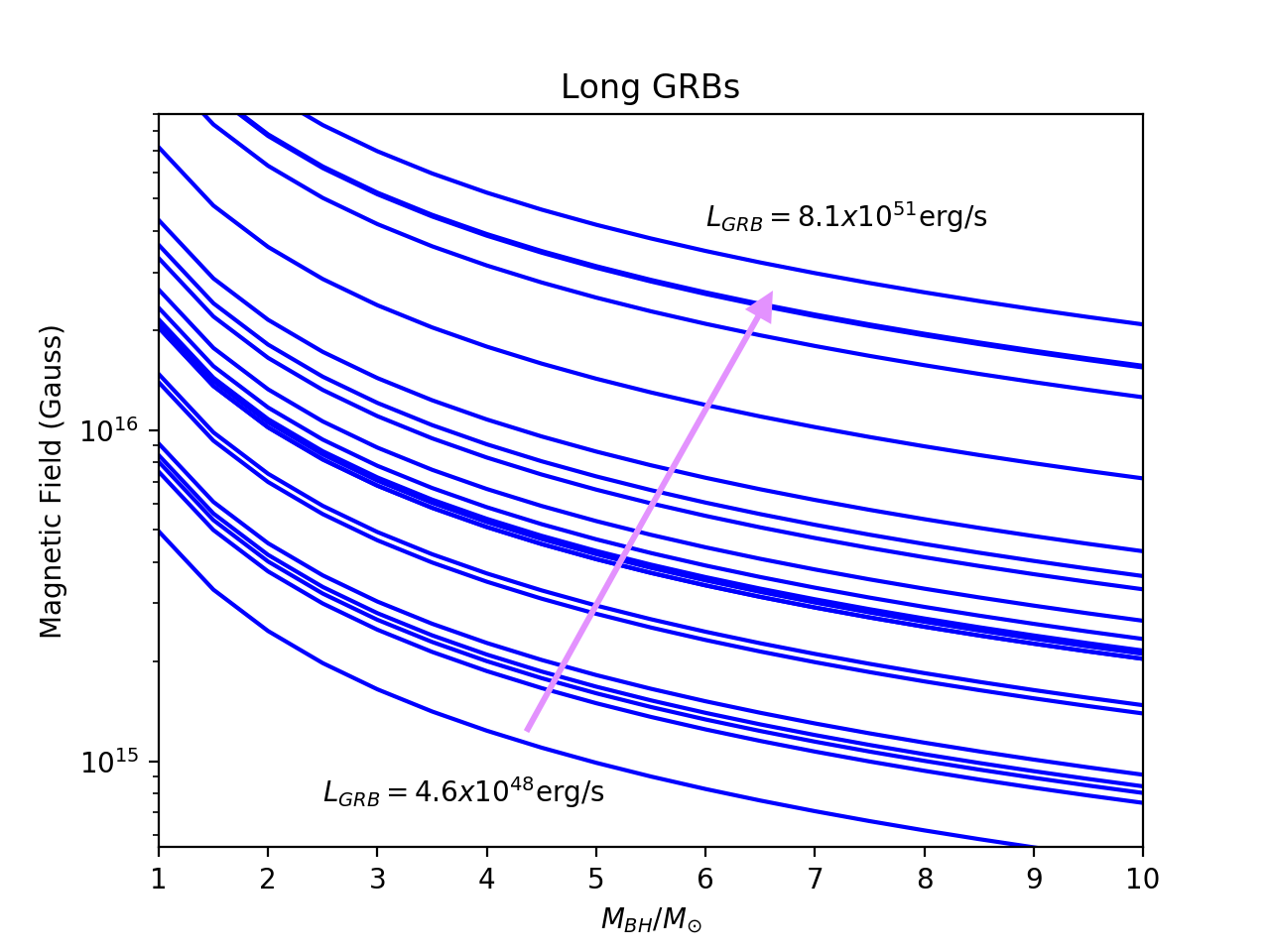}\includegraphics[width=3.5in,height=2.7in]{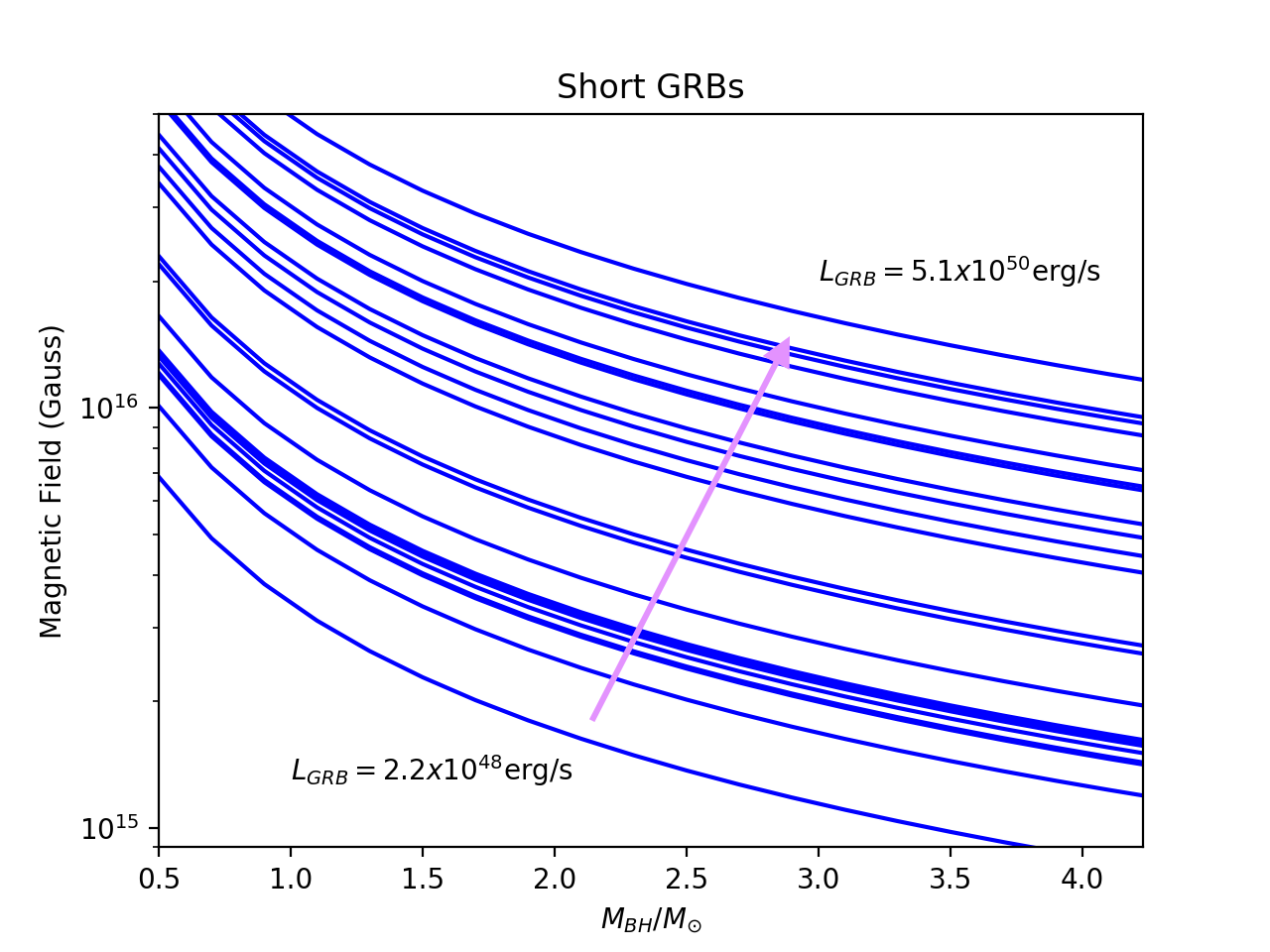}
\caption{Inferred magnetic field as a function of BH mass for 22 long GRBs (left panel) and 22 short GRBs (right panel). Each blue line represents a different observed GRB luminosity, with luminosity increasing from bottom to top.  The spin parameter of the BH is assumed to be $a=0.9$.}
\label{fig:bfmbh}
\end{figure*}
  
\section{Results}  
\subsection{Long GRBs}
  Given the assumptions above, we can explore the range of magnetic fields and BH masses required to produce observed GRB luminosities. For a representative set of long GRB luminosities, we use those in Table 1 from \cite{wu16}.  The \cite{wu16} sample is originally from \cite{sonbas15} who analyze a sample of {\em Swift} and Fermi GRBs with measurable prompt (gamma-ray) temporal statistics.  For our purposes, we consider this sample representative of the long GRB population.
  

  To estimate a reasonable range of BH masses, we assume the GRB comes from either a massive star collapse or a Helium merger (and indeed produces a BH rather than a NS inner engine). In this scenario, we can estimate the expected BH mass from the duration of the long GRB.  For example, for typical long GRB durations between $\sim 20 - 100$ s, accreting at $0.1 M_{\odot}/s$ \citep{PWF99}, we need at least $2$ to $10 M_{\odot}$ of material in the disk.  For the progenitors mentioned above, this leaves anywhere from a few to tens of solar masses left over to form the BH (ignoring any additional mass expelled in the process; note the black hole seed mass from the iron core is $\sim 2 M_{\odot}$ for a $10 - 15 M_{\odot}$ star before collapse; \cite{WH06}).  

  Within this mass range, we can then compute the magnetic field needed to satisfy equation 1, given our observed GRB luminosities.  The left panel of Figure~\ref{fig:bfmbh} shows the magnetic field at the horizon as a function of BH mass, for 22 values of observed long GRB luminosities.
As one can see in this figure, for BH masses between $2$ and $10$ solar masses, we require field strengths from $\sim 5x10^{14} G$ to $\sim 10^{17} G$ to power the GRB jet.  We discuss both the generation of these magnetic fields as well as the implications of such high magnetic fields in \S 4 below.

\subsubsection{Progenitor Magnetic Field}
  If we naively assume all of our BH flux originated from the progenitor star of the GRB, we can make an estimate of the nascent magnetic field needed to provide our required flux on the BH. In other words, we can get a crude estimate of the progenitor (``p'') magnetic field under the simple assumption of magnetic flux conservation $B_{p}R_{p}^{2} \approx B_{BH}R_{BH}^{2}$ (where we have assumed flux conservation over the same solid angle between the progenitor and BH system).  
  
\begin{figure*}
\centering
\includegraphics[width=3.5in,height=2.7in]{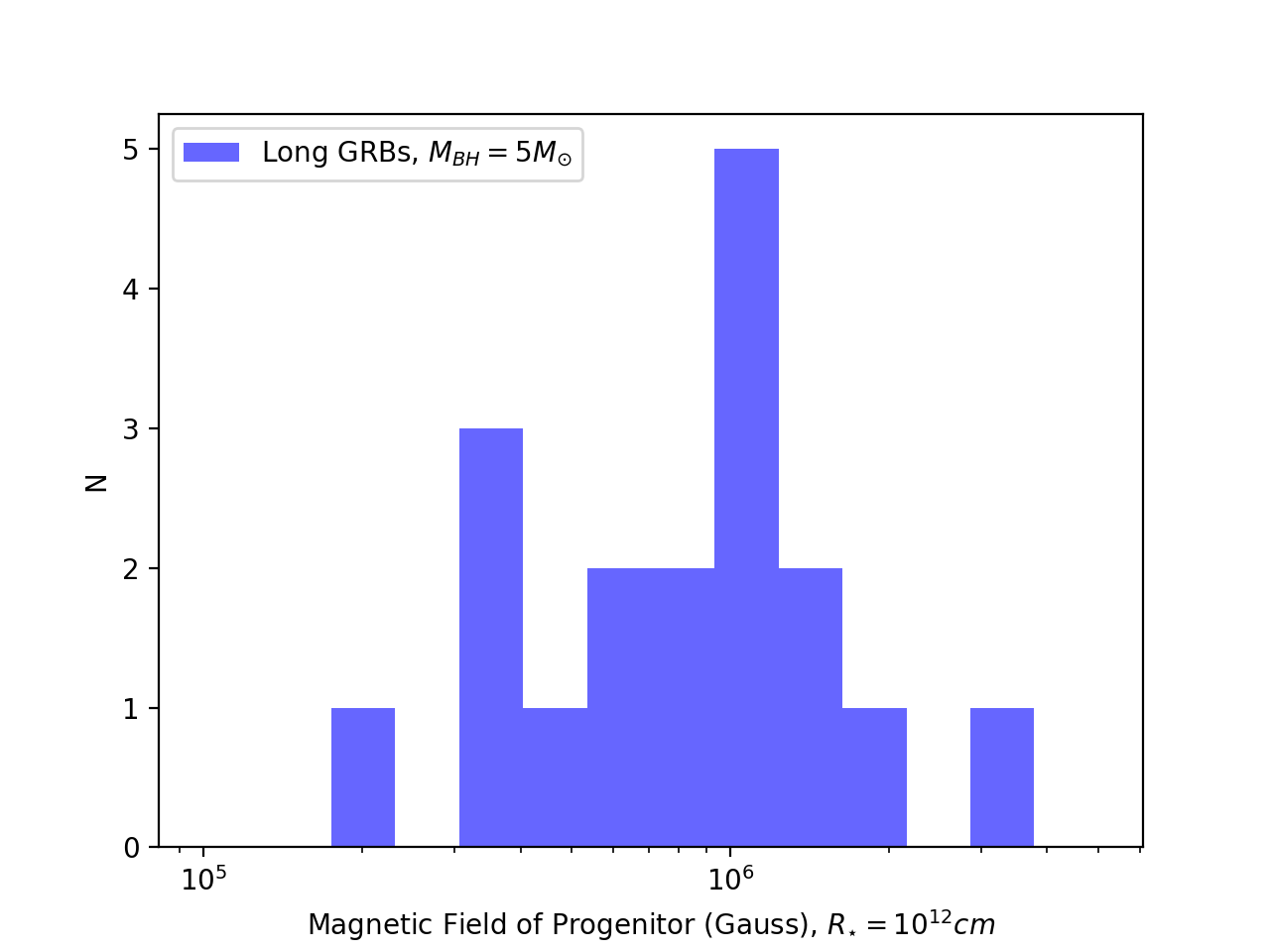}\includegraphics[width=3.5in,height=2.7in]{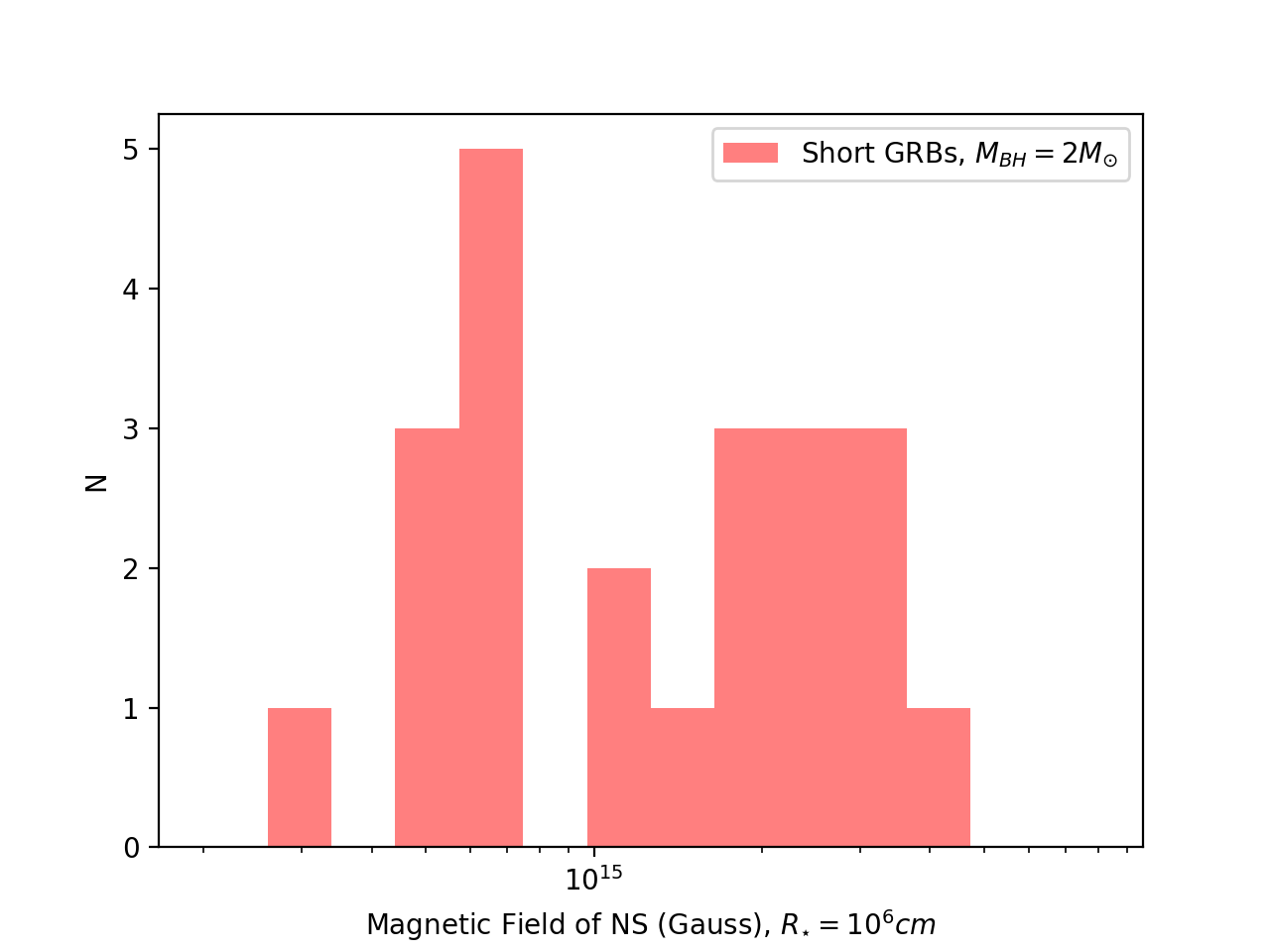}
\caption{Inferred magnetic field of GRB progenitor star with radius $R_{\star} = 10^{12}$ cm, assuming magnetic flux is conserved (left panel).  Inferred magnetic field of a NS with radius $R = 10^{6}$ cm, under the assumption that the short GRB is a result of the merger of two NSs, and the magnetic flux of the two NSs was conserved and equal to the magnetic flux of the short GRB jet (right panel).}
\label{fig:bprog}
\end{figure*}

  The left panel of Figure~\ref{fig:bprog} shows a histogram of magnetic fields for a single progenitor of radius $R_{\star} = 10^{12} \rm cm$ and a remnant black hole mass of $5 M_{\odot}$, under the assumption of magnetic flux conservation. The inferred fields are roughly $\sim 10^{6}$ G.  Typical {\em surface} magnetic fields of massive stars have been observed to be anywhere from 300G to 30kG \citep{wal12,hub16}. However, we are currently unable to probe the magnetic fields in the interiors of these stars, which in principle may be much larger. Recent astroseismology analysis of intermediate mass stars ($\sim 2 M_{\odot}$) using Kepler data have shown that some stars exhibit evidence of interior fields $\sim 10^{6}$ to $10^{7}$ G \citep{stel16}.  Hence, a dynamo in the convective cores of massive stars may indeed be able to generate such fields.  Even if this is a rare phenomenon, it is possible that GRBs come from the subset of those stars able to produce such fields.

\subsection{Short GRBs}
 As mentioned above, we assume the progenitor system for short GRBs is the merger of two NSs. We take our short GRB luminosities from \cite{fong15}, where we define GRB prompt isotropic luminosity as $E_{iso}/T_{90}$.  The \cite{fong15} sample includes all short GRBs with measured redshifts, and we consider this sample representative of the short GRB population.

 Because NSs fall in a fairly narrow mass range, we have a limited mass range for a resultant BH, which we take between $0.5$ and $4.0$ $M_{\odot}$. The right panel of Figure~\ref{fig:bfmbh} shows the magnetic field at the horizon as a function of BH mass, for 22 representative values of observed short GRB luminosities. For our range of BH masses, we require field strengths from $\sim 10^{15} G$ to $\sim 10^{17} G$ to power the short GRB jet.
  
\subsubsection{Progenitor Magnetic Field}
  We can again simply naively assume that all of the magnetic flux in the BH-accretion disk system came from the progenitor system and was conserved.  Under this (admittedly overly-simplistic) assumption, we can estimate the magnetic fields of the two NSs before the merger.
The right panel of Figure~\ref{fig:bprog} shows a histogram of the inferred magnetic fields of a DNS system in which the radius of each NS is $R_{\star} = 10^{6} \rm cm$.  

  These fields span the range from $3x10^{14}$ G to $5x10^{15}$ G.  This is in the range inferred for so-called magnetars \citep{Usov92,Thomp94}, and NSs with magnetic fields in these ranges have been invoked as the engine behind a number of astrophysical phenomena, including long and short GRBs, soft gamma-ray repeaters, fast radio bursts, and superluminous supernovae (see, for example, the discussion in \cite{Mar18}).  For larger assumed NS radius (or alternatively a harder equation of state), these field strengths will decrease.


\section{On High Magnetic Fields}
   
   The magnetic fields inferred from our analysis are extreme, and it is worth discussing how such fields in practice can be generated and sustained.  A number of studies have looked at the growth of magnetic fields, particularly in the case of double NS mergers. In these systems, \cite{Ros02} showed that massive fields $\sim 10^{17} G$ can be produced through differential rotation of the central object in a DNS merger. Both \cite{PR06} and \cite{Ob10} showed that Kelvin-Helmholtz instabilities can amplify fields to $\sim 10^{16} G$ on very short timescales ($\sim ms$), although \cite{Ob10} argue these fields are not long lasting (only a few milliseconds). \cite{ZM13} examined turbulent amplification of magnetic fields in DNS systems and showed fields up to $\sim 10^{16} G$ can be produced.
  
  For the more general case of a BH-accretion disk system, it is well known that a small extant magnetic field in the disk can be amplified by the magneto-rotational instability (MRI)  \citep{V59, Chan60,AH73,BH91}.
 
\subsection{MRI Growth}
   Simulations that examine the MRI in accretion disks\footnote{See \cite{FA13} and references therein for a review.} generally start with some seed magnetic field $B_{o}$, which grows at a rate $B=B_{o}e^{\omega t}$; the parameter $\omega$ is the maximum growth rate of the MRI, $\omega = (1/2 )r d\varpi/dr$, where  $\varpi$ is the rotational velocity in the disk \citep{BH98}.
   
  
  
  From this equation for the growth rate, we can estimate the magnetic field growth from the MRI in the context of GRB disks.  Because $\varpi$ (which for our purposes here we simply take as the Keplerian velocity in the disk) depends on radius, we evaluate the maximum growth rate at some fiducial radius. One possibility for this radius is the innermost stable circular orbit (ISCO), $R_{isco} = 6GM_{BH}/c^{2}$ for a non-rotating BH. For prograde rotation, which is a reasonable assumption for the BH and the disk, since they arose from the same (rotating) progenitor, the ISCO will be smaller than this and is given by:
   \begin{equation}
   \begin{split}
   R_{isco} & = GM_{BH}/c^{2}(3 + Z_{2} + \sqrt{(3-Z_{1})(3+Z_{1}+2Z_{2})} \\
  & Z_{1} = 1 + (1-x^{2})^{1/3}[(1+x)^{1/3}+(1-x)^{1/3}]\\
  & Z_{2} = \sqrt{3x^{2}+Z_{1}^{2}}
   \end{split}
   \end{equation}
\noindent and $x = a/R_{Sch} = ac^{2}/2M_{BH}$.


  For the range of BH masses and spins we consider above for both long and short GRBs, we find the maximum growth rate $\omega \sim 10^{5} s^{-1}$.  This means for a seed field $B_{o} \sim \mu$G, the MRI will take only $t \sim 5x10^{-4}$s to reach fields of $B \sim 10^{15}$G at the maximum growth rate.  This timescale is much shorter than the relevant timescales in both long and short GRBs, $ M/\dot{M} \sim $ seconds (where, again, $M$ is the mass in the disk and $\dot{M}$ is the accretion rate). In other words, if the MRI operates in a GRB disk in the linear regime, it is an efficient way to grow the magnetic field to the values we require to power the GRB-BZ jet.
  
   This is of course a simple analytic estimate and ignores the non-linearities that come into play as the field develops.  In particular, we've not considered at what point the magnetic field growth may saturate in the GRB disk.  This is not a well understood problem and depends on many factors related to the microphysics as well as global structure of the disk.  \cite{PWF99} estimated GRB magnetic fields in a BZ context, assuming that the magnetic field energy density reaches $1\%$ of the accretion disk kinetic energy density: $B^{2}/8\pi \sim 0.01\rho v^{2}$, where $\rho$ is the density in the disk and $v$ is the velocity.  Using an $\alpha-$ disk prescription \citep{SS73}, they estimated magnetic fields in the range of $10^{14} - 10^{16} G$ (see their Table 4).  Allowing the magnetic field energy to reach closer to equipartition with the disk energy can allow fields up to $10^{18} G$, under these arguments.

   We are also neglecting that the field grows at a different rate in different parts of the disk (the growth rate is fastest at the ISCO), and that the orbital velocity is not Keplerian near the horizon. Numerical simulations (see, e.g., figure 7 in \cite{SDGN12}) indicate that magnetically-driven turbulence becomes dominant in the disk, and thus that the magnetic field is large, after about 500 $GM_{BH}/c^3$, or $t\sim 5\times 10^{-3}$ seconds. This is an order of magnitude longer than our rough growth time estimate above, but still far shorter than the characteristic time of the GRB.  For further discussions of these issues, we refer the reader to \cite{PP05,BP07,BS13} and references therein.

\subsection{MAD Disks}
  Given the possibility of growth of a large amount of magnetic flux near the ISCO of our BH-accretion disk, our system may very well be in a magnetically arrested disk (MAD) configuration \citep{NIA03}.  In a MAD state, the magnetic field pressure is large enough to balance the ram pressure of the infalling matter, and the accretion can be halted or arrested.  Although it is an open question whether the disk can actually maintain such a high flux near the BH, simulations seem to indicate it may be viable (see, e.g., \cite{tch11} who examined MAD disks for similar values of magnetic flux and spin that we consider here).  Studies by \cite{Lis18,LTQ18} and \cite{TAM18} found that even thin disks can maintain large magnetic flux on rotating BHs and a powerful outflow is launched (note these simulations were done in the context of AGN disks).  
  
  Recently \cite{lr16,LR18} examined GRB properties in the context of the MAD model and found that the GRB variability timescale, as well as observed correlations \citep{wu16}  between: {\bf a)} the bulk Lorentz factor $\Gamma$ with minimum variability timescale $t_{min}$ in the gamma-ray light curve ($t_{min} \propto \Gamma^{-4.8 \pm 1.5}$ ), and {\bf b)} gamma-ray luminosity with the minimum variability timescale in the prompt light curve ($t_{min} \propto L^{-1.1 \pm 0.1 }$) are naturally explained in a MAD model with a Blanford-Znajek jet.  
  
  In \cite{LR18}, accretion rates of long GRBs were estimated in the context of the MAD model using
  \begin{equation}
 \dot{M} = (0.1 M_{\odot} s^{-1}) a^{-2}\eta^{-1}\epsilon_{MAD} (L_{GRB}/10^{52}erg s^{-1})
 \end{equation}

\noindent where $\epsilon_{MAD} \sim 0.001$ is a parameter estimating the degree of ``arrestedness'' of the accretion flow in the disc.  Using isotropic equivalent luminosities, they found accretion rates for long GRBs $\sim 0.5 M_{\odot}/s$, although ranging from $\sim 0.005$ to $10 M_{\odot}$ in the most extreme cases (see their Table 1). Because the average luminosity of our sample of short GRBs is lower than that of the long GRBs by roughly an order of magnitude, the accretion rates for short GRBs will be correspondingly lower by the same factor (i.e. $\dot{M} \sim 0.05 M_{\odot}/s$). Correcting for the opening angle of the jet, these rates would drop by a factor of $\sim 0.1$ (our assumed jet opening angle in this work).  \\

Finally, we briefly comment on QED processes that arise in the presence of strong magnetic fields. When the energy associated with the gyro-frequency of an electron ($\omega_{g}$) is equal to twice its rest mass ($m_{e}$) energy, pair production can occur in the presence of a magnetic field.  This requires $h \omega_{g} \gtrsim 2m_{e}c^{2}$, where $\omega_{g} = eB/m_{e}c$ . Therefore in fields greater than $B \gtrsim 2m_{e}^{2}c^{3}/h \approx 4 x 10^{13} G$, electron-positron pairs can be produced \citep{DH83}.  Some amount of pair production is necessary to keep the region around the BH force-free which allows the BZ process to operate with maximum power output \citep{MT82}.  However, a third order QED process - photon splitting - can suppress pair production in high magnetic field (this has long been suggested as a mechanism to explain radio quiet pulsars \cite{BH95,BH01}), if all three channels of photon splitting operate \citep{BH01}.  

  In this work, we assume that neither of these processes (pair production or photon splitting) measurably compare to the energy in our magnetic field, nor do they affect the operation of the BZ process, so that they can effectively be ignored.   A detailed examination of these effects in very strong magnetic fields is the subject of a future investigation.



\section{Progenitor Angular Momentum}
  Throughout this paper, we have assumed a BH spin parameter of $a \sim 0.9$.  As discussed above and displayed in Figure~\ref{fig:bhspins}, there is ample observational evidence for BH spin parameters of this magnitude, in both stellar mass and supermassive BHs.  Theoretically, it is still a question how to achieve these spin parameters in the process of making the BH. Below, we discuss the conditions for which that assumption is reasonable, given our long and short gamma-ray progenitors. 

\begin{figure}
\centering
\includegraphics[width=\columnwidth, height=2.7in]{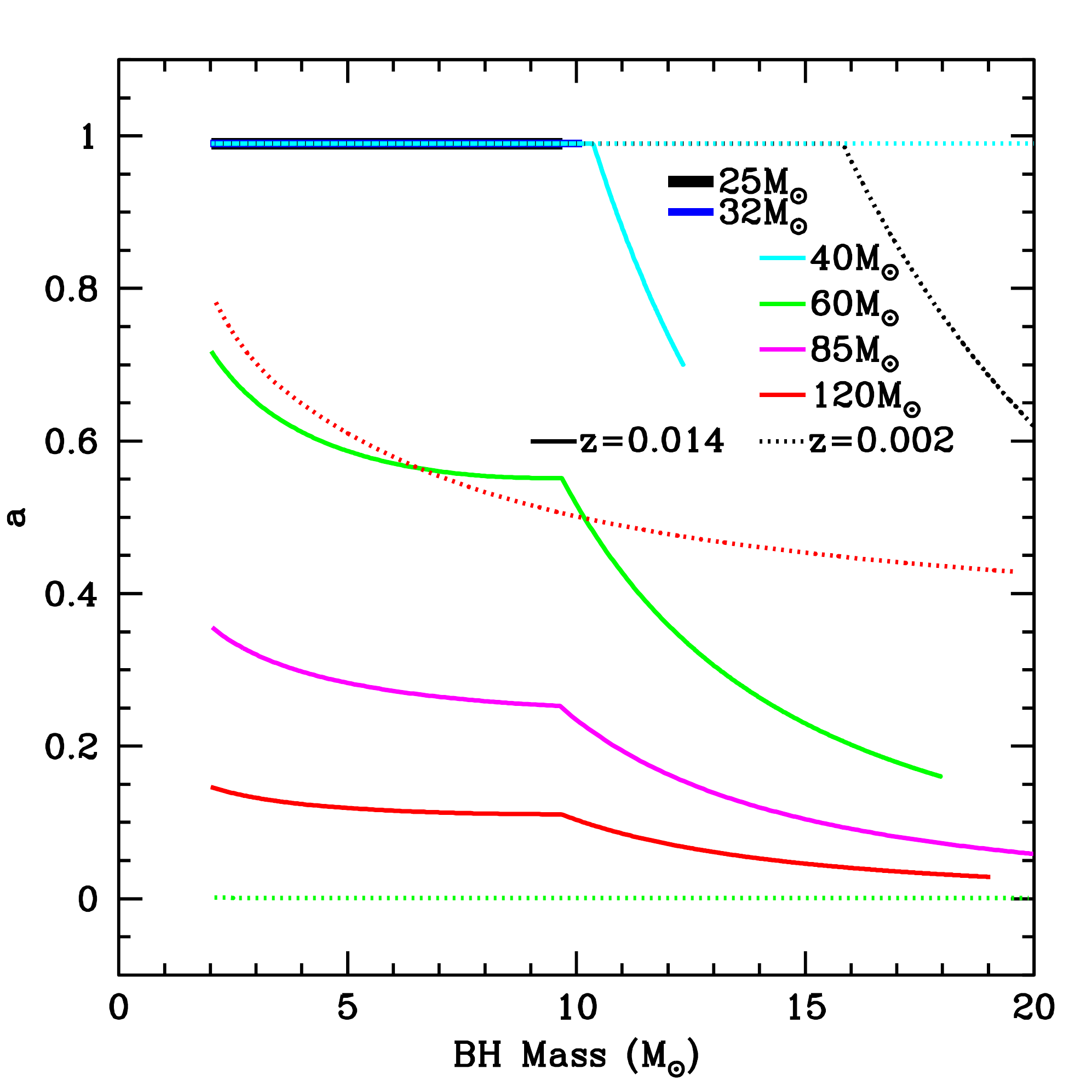}
\caption{BH spin parameters as a function of core mass for stellar models over a range of zero-age main sequence mass and two metallicities~\citep{belczynski17}. }
\label{fig:avm}
\end{figure} 

\subsection{Long GRBs}

Figure~\ref{fig:avm} shows the spin parameter as a function of BH mass for a variety of initial stellar masses and metallicities, using a suite of models from \cite{belczynski17}, which employs the {\em Geneva} stellar evolutionary code~\citep{Geo13}.  These models assume there are no dynamo-generated magnetic fields driving coupling between layers in the stars, and give a maximum to the rotation rate of any single star.  Material is accreted onto the BH layer by layer and the angular momentum added to the BH is set by the angular momentum of the accreting material:  $j=min(j_{\rm infall},j_{\rm ISCO}$) where $j_{\rm infall}$ is the specific angular momentum of the accreting material and $j_{\rm ISCO}$ is the angular momentum of a disk at the innermost stable circular orbit.  This equation assumes that, if there is sufficient angular momentum to form a disk, only the angular momentum at the innermost stable circular orbit is accreted onto the BH.  From this process, we can estimate the spin of the BH.   

\begin{figure}
\centering
\includegraphics[width=\columnwidth, height=2.7in]{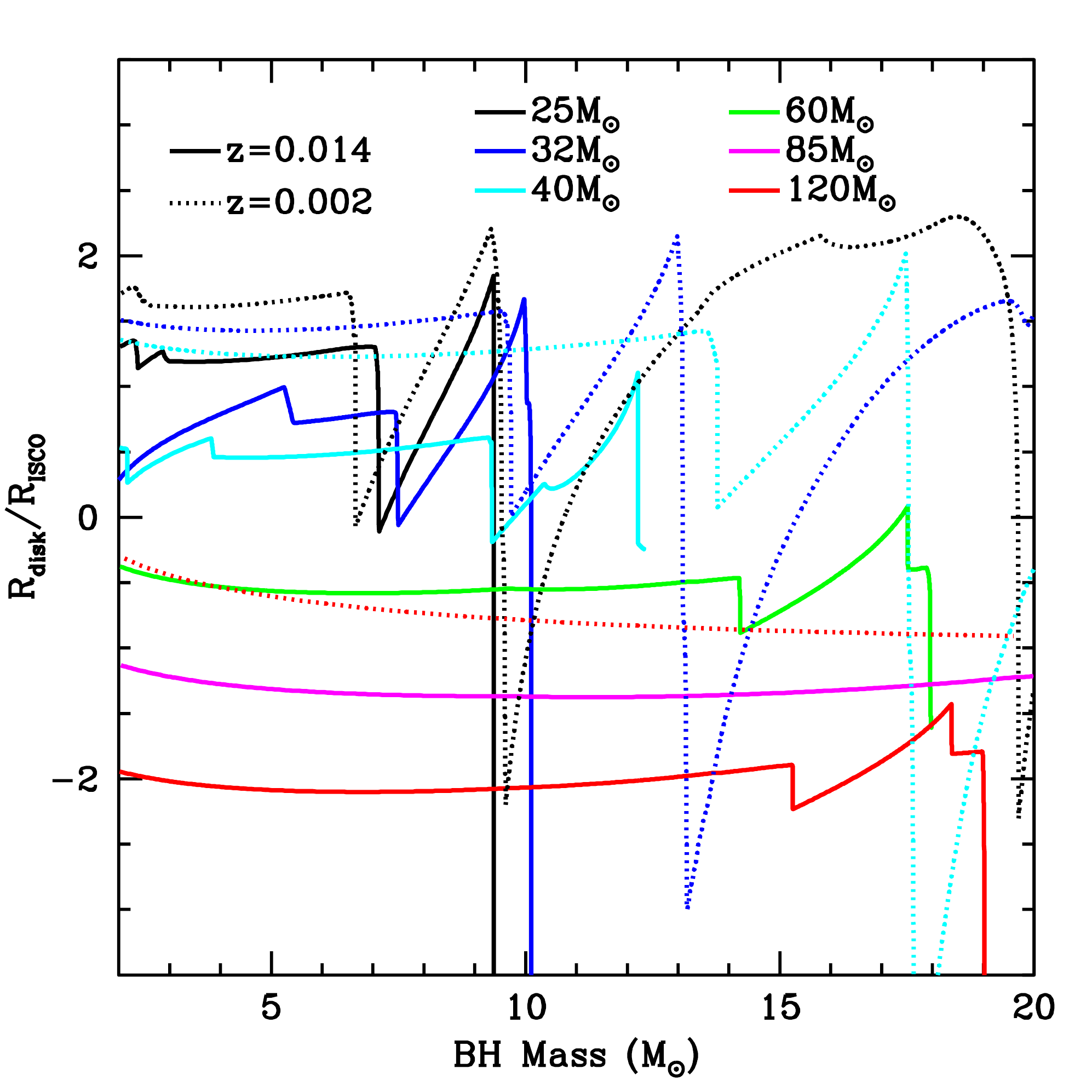}
\caption{Radius of the disk formed in the collapsing star as a function of the interior mass for the models discussed in Figure~\ref{fig:avm}.  For our slowly-rotating progenitors, this disk or hang-up radius is less than the innermost stable circular orbit, meaning no stable disk is formed.}
\label{fig:rdisk}
\end{figure}

  A wide range of angular momenta can be produced in stellar models, depending both on mass and metallicity.  Not all of these systems will actually form the BH accretion disks needed to form GRBs.  The size of the disk produced in these collapsing stars is determined by the radius at which the centrifugal force of the rotating, infalling material equals the gravitational force (based on the enclosed mass).  For many of these systems, there is not enough angular momentum and this radius falls below the ISCO (so that no disk forms).  Some systems with enough angular momentum (corresponding to a BH spin parameter of $\sim 0.8$) do form a disk, depending on the initial properties of the progenitor and the BH mass. Figure~\ref{fig:rdisk} illustrates this point. This plot finds the radius where the centrifugal force equals gravity:  $r_{\rm disk}=j^2/(G M_{\rm BH})$ where $j$ is the specific angular momentum of the infalling material, $G$ is the gravitational constant and $M_{\rm BH}$ is the mass of the BH set by the mass of the material enclosed at each Lagrangian mass point.  
  
  A number of mechanisms have been studied showing that angular momentum is transported out of stellar cores, producing even more slowly rotating cores (see, e.g., the recent review by \cite{AMR18}); these stars will not have enough angular momentum to produce GRBs \citep{JSP18}.  Hence, GRBs may be the special subset of collapsing stars for which the remnant BH spin parameter is high (above 0.9).
  
 
The helium merger model (produced when a NS or BH spirals into the helium core of a massive star) produces even faster rotation rates~\citep{fryer98,zhang01}.  The disks formed tend to be large (above 10,000\,km) and the subsequent accretion rates are lower.  The BH spins for all of these systems will quickly spin up to values above 0.8-0.9 unless the initial compact remnant mass is large compared to the helium core.


\subsection{Short GRBs}
   The spin period of the remnant of a double NS  merger is very much an open question and depends on a number of things, including the masses of the NSs, their equations of state, the initial spins of each NS, etc. (for a recent discussion of these issues, see \cite{Rad16,PGP17}). Using results of Newtonian merger calculations \citep{kor12} and various equation of state studies, \cite{Fry15} examined a suite of models to determine the fate of the compact remnant of a DNS merger (see, e.g. their Table 1).  For core masses above $2 M_{\odot}$, they find angular momenta (column 5 of their Table 1) corresponding to remnant core spin parameters between $\sim 0.35$ and $\sim 0.65$. \cite{Zappa18} looked at the spin parameter of the BH remnant of a DNS merger using a large sample of numerical relativity simulations with different binary parameters and input physics. They find BH remnants with spin parameters between $0.6$ and $0.9$ for a wide range of models (finding an empirical relation between the the total gravitational radiation and the angular momentum of the remnant; see their Figure 4). Finally, \cite{Ruiz16} also simulated DNS mergers for equal mass, magnetized NSs with an n=1 polytrope equation of state.  They find a BH remnant with $a \sim 0.74$. These initial spins from the BH remnant of a DNS merger approach the high value of the spin parameter we assumed above to explain short GRB luminosities.  
   
 
\section{Conclusions}
  We have examined the constraints that observed GRB luminosities place on GRB inner engine properties in the context of a BZ jet powering the GRB.  Our main results are as follows:
  \begin{itemize}
\item{For long GRBs with BH masses in the range of $2 - 10 M_{\odot}$, and a BH spin parameter of $a \sim 0.9$, magnetic fields from $\sim 10^{17} G$ (for the less massive BHs) down to $\sim 5 x 10^{15} G$ (for more massive BHs) are needed to explain the range of observed GRB luminosities, assuming $10 \%$ of the BZ power goes into GRB emission.}
\item{For short GRBs with BH masses in the rage of $0.5 - 4 M_{\odot}$, and a BH spin parameter of $a \sim 0.9$, magnetic fields from $\sim 10^{17} G$ (for the less massive BHs) down to $\sim 10^{15} G$ (for more massive BHs) are needed to explain the observed sGRB luminosity assuming $10 \%$ efficiency.}
\item{The inferred fields of the progenitor systems under the simple assumption of magnetic flux conservation are $\sim 10^{6}$ G for single star/collapsar systems powering long GRBs, and $\sim 10^{15} G$ for double NS systems powering short GRBs.}
\item{These magnetic field values can be reached through MRI growth on timescales much shorter than the duration or variability timescales in GRBs.  The consequences of such high magnetic flux include magnetically arrested disks (MADs), which have been shown to explain a number of observed phenomena in GRB light curves \citep{lr16,LR18}.}
\end{itemize}

We have assumed a relatively conservative jet opening angle of $10^{\circ}$ throughout.  For more narrowly beamed systems (and therefore less emitted luminosity), the constraints on our BZ jets are lessened. 

  The detailed properties of GRB disks and how they ultimately relate to the BZ jet power need to be explored further. For example, the estimates in \cite{PWF99} indicate that for a constant accretion rate, the disk density will scale inversely with BH mass; therefore - under the assumption that the magnetic field is some fraction of the disk kinetic energy - higher mass BHs will result in smaller magnetic fields.  This simple scaling argument, however, should be examined in detail.  In particular, the physics of the accretion process and how the rate of accretion scales with BH mass (and varies throughout a GRB) are important processes that will affect this picture.  
  We note that that in this work, we have neglected any contribution from a jet powered by neutrino annihilation \citep{Eichler89}. For some disk models, this process could enhance jet power, possibly substantially.

 Finally, simulations exploring the generation of high magnetic flux at the BH horizon and its consequences, as well as a more detailed look at the connection between progenitor angular momentum and the spin parameter of the BH remnant could help validate some of the assumptions in this work, and - more importantly - uncover the nature of the GRB inner engine.

\section*{Acknowledgements}
  We are indebted to the referee for a very careful reading of this manuscript and numerous valuable suggestions. N. L.-R. thanks Peter Polko and Oleg Korobkin for interesting discussions on accretion disks and DNS mergers, respectively, Greg Salvesen for discussions and references on BH spins, and Thomas Maccarone for bringing up the issue of photon splitting in strong magnetic fields. C.T. thanks the NSF Step program for financial support while some of this work was carried out. This work was supported by the US Department of Energy through the Los Alamos National Laboratory. Additional funding was provided by the Laboratory Directed Research and Development Program and the Center for Nonlinear Studies at Los Alamos National Laboratory under project number 20170508DR. Los Alamos National Laboratory is operated by Triad National Security, LLC, for the National Nuclear Security Administration of U.S. Department of Energy (Contract No. 89233218CNA000001). LA-UR-18-31627





\bibliographystyle{mnras}
\bibliography{refs}







\bsp	
\label{lastpage}
\end{document}